\documentclass[aps,prl,twocolumn,superscriptaddress]{revtex4-1}

\usepackage{subfigure}
\usepackage{graphicx}
\def\dt{\delta\tau}		

\def\hats{\hat S}		
\def\hatt{\hat T}		
\def\SST{\{S,\{S,T\}\}}
\def\TST{\{T,\{S,T\}\}}
\def\TTTST{\{T,\{T,\{T,\{S,T\}\}\}\}}
\def\STSST{\{\{S,T\},\{S,\{S,T\}\}\}}
\def\TSSST{\{T,\{S,\{S,\{S,T\}\}\}\}}
\def\TTSST{\{T,\{T,\{S,\{S,T\}\}\}\}}
\def\SSSST{\{S,\{S,\{S,\{S,T\}\}\}\}}
\def\STTST{\{\{S,T\},\{T,\{S,T\}\}\}}
\def\O{{\cal O}}		
\def\Var{\mathop{\rm Var}}      
\def\erfc{\mathop{\rm erfc}}    
\def\half{\frac12}
\def\acc{{\hbox{\tiny acc}}}
\def\PQPQP{{\hbox{\tiny PQPQP}}}
\def\FG{{\hbox{\tiny FG}}}
\begin{document}


\title{
Improving dynamical lattice QCD simulations through integrator
tuning using Poisson brackets and a force-gradient integrator}



\author{M. A. Clark}
\affiliation{Harvard-Smithsonian Center for Astrophysics, 
  Cambridge, MA 02138, U.S.A.}

\author{B\'alint Jo\'o} 
\affiliation{Jefferson Lab, 12000 Jefferson Avenue, Newport News, VA 23606,
  U.S.A.}

\author{A. D. Kennedy}  
\affiliation{Tait Institute and SUPA, School of Physics \& Astronomy, The
  University of Edinburgh, Edinburgh EH9 3JZ, Scotland, U.K.}

\author{P. J. Silva}
\email[Corresponding author: ]{psilva@teor.fis.uc.pt}
\affiliation{Centro de F\'isica Computacional, Universidade de Coimbra,
  Portugal}


\date{August 8, 2011}

\begin{abstract}
We show how the integrators used for the molecular dynamics step 
of the Hybrid Monte Carlo algorithm can be further improved. 
These integrators not only approximately conserve some Hamiltonian $H$ 
but conserve exactly a nearby shadow Hamiltonian $\tilde{H}$. This property 
allows for a new tuning method of the molecular dynamics 
integrator and also allows for a new class of integrators (force-gradient 
integrators) which is expected to reduce significantly the computational 
cost of future large-scale gauge field ensemble generation.
\end{abstract}


\maketitle

\section{Introduction and motivation}

Hybrid Monte Carlo (HMC) \cite{hmc} is the algorithm of choice to generate
lattice QCD configurations including the effect of dynamical fermions.
The most time consuming ingredient of HMC is the molecular dynamics (MD) step,
which consists of a reversible volume-preserving approximate MD trajectory of
\(\tau/\dt\) steps (with \(\tau\) being the length of the trajectory and
\(\dt\) the stepsize) followed by a Metropolis accept/reject test with
acceptance probability \(\min(1,e^{-\delta H})\) where \(\delta H\) is the
change in the Hamiltonian \(H=T+S\) whose kinetic and potential parts
are \(T\) and \(S\).

A molecular dynamics trajectory is not only an approximate integral curve of
the Hamiltonian vector field \(\hat H\) corresponding to \(H\), but is also an
exact integral curve of the Hamiltonian vector field \(\widehat{\tilde H}\) of
an exactly conserved shadow Hamiltonian~\(\tilde H\).  The asymptotic expansion
of this shadow Hamiltonian in the stepsize \(\dt\) may be computed using the
Baker--Campbell--Hausdorff (BCH) formula and expressed in terms of Poisson
brackets (PBs)~\cite{inexact, lat07}.
As a simple example consider the PQPQP (also known 
as 2MN \cite{omelyan}) integrator
\begin{displaymath}
  U_\PQPQP(\tau) = \left(e^{\lambda\hats\dt} e^{\half\hatt\dt}
  e^{(1-2\lambda)\hats\dt} e^{\half\hatt\dt} e^{\lambda\hats\dt} \right)^{\tau/\dt}
\label{pqpqp}
\end{displaymath}
whose shadow Hamiltonian is 
\begin{widetext}
\begin{eqnarray}
\tilde H_{PQPQP} &=& H +
\left(\frac{6\lambda^2 -6\lambda +1}{12} \SST + \frac{1-6\lambda}{24}
\TST\right)\dt^2 \\ 
&+& \biggl(
\frac{-1+30\,{\lambda}^{2}-60\,{\lambda}^{3}+30\,{\lambda}^{4}}{720} \SSSST
+ \frac{-4+15\,\lambda+15\,{\lambda}^{2}-30\,{\lambda}^{3}}{720} \TSSST
\nonumber \\
&+& \frac{-7+30\lambda}{1440} \TTSST + \frac{-7+30\lambda}{5760} \TTTST
\nonumber \\
&+& \frac{-2+15\,\lambda-35\,{\lambda}^{2}+30\,{\lambda}^{3}}{240} \STSST  
+ \frac{-2+15\,\lambda-30\,{\lambda}^{2} }{720} \STTST    \biggr)\dt^4
+ \O(\dt^6).\nonumber
\label{shadow:pqpqp}
\end{eqnarray}
\end{widetext}
Note that we have one free parameter, \(\lambda\), which is
often set to some value not taking PBs into account. In \cite{forcrand}, 
the authors  chose \(\lambda\) by minimising  \(\delta H\) empirically, 
requiring a sequence of runs at different values of \(\lambda\). 
Others used $\lambda_c\approx 0.193183$ \cite{omelyan}, which minimizes 
the norm of the coefficients of the PBs in the second-order term. 
However, this is not necessarily the best choice.

We have evaluated PBs and shadow Hamiltonians for gauge theories (where
gauge fields are constrained to live on a Lie group manifold) for the
first time \cite{lat07, lat08, lat09, lat10}.  Therefore, in this
letter we propose to measure the volume-averaged PBs and 
tune the free parameters of an
MD integrator taking PB measurements into account. As we will see, our
tuning procedure also allows us to find out the best number of steps
of a nested integrator scheme. We also present a new integrator step and a 
new integrator which will be able to reduce the cost of large volume 
simulations.
\vspace*{-0.2cm}
\section{Integrator tuning}
\vspace*{-0.1cm}
\begin{figure}[!t] 
  \includegraphics[origin=c,angle=0,width=8cm]{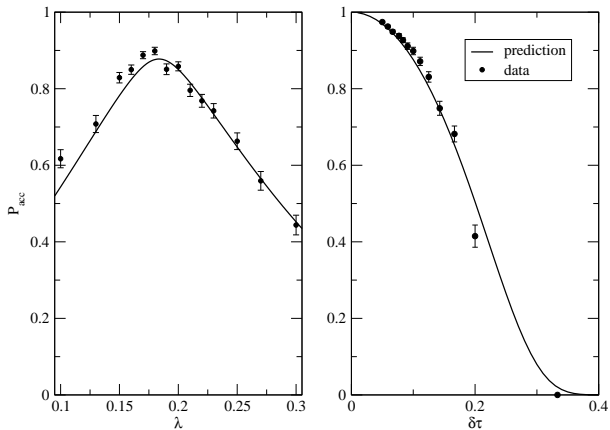}
\caption{Comparison of measured acceptance rates and
their predictions from Poisson bracket measurements. 
In the left hand plot we fix \(\dt=0.1\) and leave \(\lambda\) 
as a free parameter, whereas in the right  we take \(\lambda=0.18\) 
and plot \(P_\acc\) as a function of $\dt$.}
   \label{tuning}
\end{figure}
Let us define the difference between the shadow (\(\tilde H\)) and actual
(\(H\)) Hamiltonians as \(\Delta H=\tilde H-H\). 
Noting that $\Var(\Delta H)$ means the variance of the distribution of
values of \(\Delta H\) over phase space, one can show that the 
acceptance rate $P_\acc$ can be given by \cite{lat10}
\begin{equation}
  P_\acc   = \erfc\left(\sqrt{\frac14\Var(\Delta H)}\right).
  \label{paccvar}
\end{equation}
To estimate \(P_\acc\) from eq. \ref{paccvar}, one only needs to
measure the PB from equilibrated configurations.  This allow us to
express \(P_\acc\) as a function of the integrator parameters and find
their optimal values that maximize \(P_\acc\).

As a simple test, we consider an HMC simulation of two flavors of
Wilson fermions at \(\kappa=0.158\) and Wilson gauge action at
\(\beta=5.6\) on an \(8^4\) lattice.  We use a single level PQPQP
integrator and a unit trajectory length, therefore we have two tunable
parameters: the integrator parameter \(\lambda\) and the step
size \(\dt\).  We measure $\Delta H$ up to fourth order in $\dt$.  In
Figure~\ref{tuning} we compare the acceptance rates predicted by the
formula above with numerical data taken from simulations at various
values of \(\lambda\) and \(\dt\).  The PB values used for the
predictions were measured at \(\lambda=0.18\) and \(\dt = 0.1\)
-- but one should note that our predictions are independent of the
integrator parameters used to get the Poisson bracket values.

The plots show good agreement between predicted and measured
acceptance rates, provided the stepsize is not too big, otherwise the
BCH expansion breaks down.  Moreover, the maximum of the acceptance
rate in the left hand plot is achieved at $\lambda_{max}\approx 0.1836$
(to be compared with $\lambda_c$).  We now use eq.  (\ref{paccvar}) to
tune the MD integrator on a larger volume. Ultimately, we are
interested in reducing the computational cost, which depends on the
wall-clock time spent computing the force terms on a unit of MD time
as well as the acceptance rate, and the 
autocorrelation time $\tau_{corr}$ for the
observables.  We neglect $\tau_{corr}$ in this discussion
as it is not sensitive to the choice of integrator parameters as
long as the acceptance rate is reasonable, and define our cost metric as
\begin{equation}
  \mbox{cost} = \frac{\mbox{trajectory CPU time}}{P_\acc  \,\, \tau}\;.
\label{cost}
\end{equation}
For our purposes, the numerator of eq. (\ref{cost}) is estimated by
considering the time spent in force computation along the
trajectory. In particular, for a nested integrator, the numerator of
this cost function is a function of the number of steps at each level
times the CPU time required to compute the forces at that
level. Therefore, minimizing eq. (\ref{cost}) will allow us to both find out
the optimal integrator parameters, as well as find out the optimal
stepsize, or the number of steps at each level of a nested integrator
scheme. This is a more direct approach than the popular ``balancing
forces'' method~\cite{urbach}.

\section{Tuning a real simulation}

\begin{table}[!h]
  \begin{center}
    \begin{tabular}{|c|c|c|c|c|}
      \hline 
      Level \(i\) & Force & F time & FG time \\
      \hline
      0 & Hasenbusch (\(\mu=0\) / \(\mu=0.057\)) & 21.21 s  & 26.61 s  \\
      1 & Hasenbusch (\(\mu=0.057\) / \(\mu=0.25\)) & 3.98 s & 7.55  s \\
      2 & Wilson (\(\mu=0.25\)) & 1.05 s & 1.98 s \\
      3 & Gauge & 0.075 s & 0.142 s \\
      \hline
    \end{tabular}
  \end{center}
  \caption[Integrator set-up]{Set-up used in the HMC simulation described in
    this section, together with typical times spent on force computation. 
    $\mu$ is the twisted mass parameter \cite{urbach}. For
    convenience, times for the force-gradient computation are also shown here.}
  \label{setup}
\end{table}

\begin{table*}[!t]
    \begin{tabular}{|c|cccc|cccc|c|c|c|c|c|}
      \hline
   Tuning   & \multicolumn{4}{|c|}{} & \multicolumn{4}{|c|}{} & 
      \multicolumn{2}{|c|}{Prediction} & \multicolumn{3}{|c|}{Measurement} \\ 
      Scheme & \multicolumn{4}{|c|}{\(m_i\)} &
      \multicolumn{4}{|c|}{\(\lambda_i\)} && \(F\) time & & Time & \\
      & 0 & 1 & 2 & 3 & 0 & 1 & 2 & 3 & \(P_\acc\) & / traj. &
      \(P_\acc\) & / traj. & Cost \\

      \hline
      Original & 3 & 1 & 2 & 3 & \(1/6\) & \(1/6\) & \(1/6\) & \(1/6\) &
      \(0.85(1)(3)\) & 308 s & \(0.88(4)\) & 405 s & 463 \\
      \hline

      PQ4 / 4th & 3 & 1 & 2 & 1 & \(0.1903(36)\) & \(0.1696(66)\) & \(0.1885(69)\) &
      \(0.1670(80)\) &
      \(0.89(1)(0)\) & 294 s & \(0.85(3)\) & 399 s & 471 \\

      PQ4 / 4th & 3 & 1 & 1 & 2 & \(0.1966(64)\) & \(0.1660(131)\) & \(0.1885(35)\) &
      \(0.1524(168)\) &
      \(0.80(2)(0)\) & 267 s & \(0.82(5)\) & 360 s & 438 \\
      \hline

      PQ3 / 4th & 3 & 3 & 2 & $-$ & \(0.1803(22)\) & \(0.1902(53)\) & \(0.1281(220)\) & --- 
      & \(0.83(3)(0)\) & 234 s & \(0.83(4)\) & 345 s & 417 \\

      PQ3 / 2nd & 3 & 3 & 2 & $-$ & \(0.1735(25)\) & \(0.1924(53)\) & \(0.1415(216)\) & --- 
      & \(0.81(2)(3)\) & 234 s & \(0.72(4)\) & 354 s & 491 \\

      PQ3       & 3 & 3 & 2 & $-$ &  \(\lambda_c\) & \(\lambda_c\) &
      \(\lambda_c\)   & ---
      & \(0.76(3)(3)\) & 234 s & \(0.80(3)\) & 339 s & 425 \\  

      PQ3      & 3 & 3 & 2 & $-$ &  \(1/6\) & \(1/6\) & \(1/6\)   & --- 
      & \(0.73(3)(5)\) & 234 s & \(0.74(5)\) & 349 s & 473 \\  

      PQ3 / 4th & 3 & 3 & 1 & $-$ & \(0.1793(24)\) & \(0.1890(65)\) & \(0.1636(66)\) & --- 
      & \(0.81(2)(0)\) & 228 s & \(0.78(5)\) & 342 s & 442 \\

      \hline
    \end{tabular}
    \caption[PQPQP tuning]{Tuning of the PQPQP integrator scheme. All errors shown are statistical, with the exception of the second set of errors in the predicted acceptance rates, which are some sort of systematic error, estimated from the difference of predicting acceptance rates using a shadow Hamiltonian up to $\dt^2$ or $\dt^4$.  All times refer to runs utilizing 128 cores of the Iridis cluster.}
  \label{tab-pqpqp}
\end{table*}

As an application of our tuning technology, we consider a HMC simulation of a
\(24^3\times 32\) lattice, with two flavours of Wilson fermions with
\(\kappa=0.1580\) and \(\beta=5.6\).  As in \cite{urbach}, we use a nested
PQPQP integrator scheme, with the inclusion of two Hasenbusch fields with
twisted mass fermions as ``preconditioners''. In \cite{urbach}, each nested
level of the integrator has one force term, and the free parameter of the PQPQP
integrator has been set to \(\lambda=1/6\) at all levels. In Table~\ref{setup},
for each level~\(i\) (note that \(0\) is the outermost level), we show the type
of force and its parameters, and mean values of the time spent on force and
force-gradient computation. 

In order to improve the integrator scheme used in \cite{urbach}, 
we considered two different nested schemes:
\begin{itemize}
\item[PQ4.] the original scheme, but with tuned values of \(\lambda\);
\item[PQ3.] the two Hasenbusch fields appear now at the same level (so we have
  only 3 different levels).
\end{itemize}
Table~\ref{tab-pqpqp} shows parameters which minimize the cost metric 
(\ref{cost}). In our simulations, we have fixed \(\tau=1\), unless stated 
otherwise. For each scheme (which is also described by the highest power of 
$\dt$ used to compute $\Delta H$), we show the optimal number of steps 
at each level, the optimal \(\lambda\) parameters, our predictions for the 
acceptance rate, the estimated time spent in force computation in one 
trajectory, and measurements of acceptance rates and trajectory times. 
For comparison we also show data for the original scheme \cite{urbach}.

We see that all predicted acceptance rates agree, within errors, with the 
measured ones. Furthermore, the tuning of $\lambda$'s in the PQ4 scheme 
allows a reduction of the number of steps on the inner levels, 
so the CPU time per trajectory decreases.
Moreover, the PQ3 scheme allows further improvement in cost measures. For this
scheme, we also show the performance obtained using other $\lambda$ values. 
We conclude that, for an optimal choice of integrator parameters, one is encouraged to tune the integrator using the best available approximation to $\Delta H$.

\section{Force-gradient integrator} 

Since the Poisson bracket $\SST$ does not depend on momentum
\cite{lat07}, we can evaluate the integrator step
$e^{\widehat{\SST}\dt^3}$ explicitly. If one uses $\lambda=1/6$ for
the PQPQP integrator together with this integrator step,
we eliminate all $\mathcal{O}(\dt^2)$ terms in $\Delta H$ .
We therefore define a PQPQP \textit{force-gradient}
integrator as
\begin{displaymath}
  U_{\FG}(\tau) = \left(e^{\frac{1}{6}\hats\dt} e^{\half\hatt\dt}
  e^{\frac{48\hats\dt-\widehat{\SST}\dt^3}{72}} e^{\half\hatt\dt} e^{\frac{1}{6}\hats\dt} \right)^{\tau/\dt}.
\label{fg}
\end{displaymath}
Note that the performance of this integrator has been shown to be much
better than Campostrini integrator \cite{lat09}.  This is not
surprising since the coefficients of the $\mathcal{O}(\dt^4)$ term in
the shadow Hamiltonian 
are about two orders of magnitude smaller \cite{lat07}.

In order to test the performance of this integrator, we considered the two integrator schemes defined in the last section, replacing the PQPQP integrator, at all levels, by this new one. We will
denote these new schemes as FG4 and FG3 which have 4 and 3 different
levels, respectively.  As in this case there are no tuneable
parameters, we could only vary the number of steps at each level.  In
Table~\ref{fg2} we show the best parameters we found as well as the
measured values of acceptance rates and trajectory times.

\begin{table}[!h]
  \begin{center}
    \begin{tabular}{|c|c|cccc|c|c|c|c|c|}
      \hline
      & \multicolumn{5}{|c|}{} & \multicolumn{2}{|c|}{Prediction} &
      \multicolumn{3}{|c|}{Measurement} \\ 
      Scheme & $\tau$ &\multicolumn4{|c|}{\(m_i\)} && \(F+FG\) && Time &  \\
      & & 0 & 1 & 2 & 3 & \(P_\acc\) & time / traj. & \(P_\acc\) & / traj. &
      cost \\
      \hline
      FG4 & 1.0 & 3 & 1 & 1 & 1 & \(0.97(1)\) & 415 s & \(0.92(2)\) & 523 s  & 569 \\
      FG4 & 1.1 & 3 & 1 & 1 & 1 & \(0.96(1)\) & 415 s & \(0.85(4)\) & 526 s  & 560 \\
      FG4 & 1.2 & 3 & 1 & 1 & 1 & \(0.94(1)\) & 415 s & \(0.71(6)\) & 500 s  & 585 \\
      FG3 & 1.0 & 3 & 1 & 1 & $-$ & \(0.91(1)\) & 314 s & \(0.80(3)\) & 393 s & 492 \\
      \hline
    \end{tabular}
  \end{center}
  \caption[FG tuning]{Tuning of FG4 and FG3 integrator schemes. 
Note that $P_{acc}$ predictions are not compatible with measurements. 
We expect this to be due to higher order corrections to $\Delta H$.}
  \label{fg2}
\end{table} 

A comparison between Tables \ref{tab-pqpqp} and \ref{fg2} shows that the use 
of a force-gradient integrator allows a smaller number of steps in the 
inner levels. Furthermore, comparing the 4 nested level schemes, one sees 
that the force-gradient acceptance rates are higher. 
However, the trajectory CPU times also increase (because we have to 
evaluate one more step), so the cost measures are higher. 

\section{A larger volume}

\begin{table*}[htb]
  \begin{center}
    \begin{tabular}{|c|cccc|cccc|c|c|c|c|c|}
      \hline
   Tuning   & \multicolumn{4}{|c|}{} & \multicolumn{4}{|c|}{} & 
      \multicolumn{2}{|c|}{Prediction} & \multicolumn{3}{|c|}{Measurement} \\ 
      Scheme & \multicolumn{4}{|c|}{\(m_i\)} &
      \multicolumn{4}{|c|}{\(\lambda_i\)} && \(F\) time && Time  & \\
      & 0 & 1 & 2 & 3 & 0 & 1 & 2 & 3 & \(P_\acc\) & / traj. &
      \(P_\acc\) & / traj. & Cost \\
      \hline
      PQ4 / 4th & 5 & 1 & 2 & 1 & \(0.1940(32)\) & \(0.1712(53)\) & \(0.1961(50)\) &
      \(0.1757(64)\) &
      \(0.90(1)(0)\) & 1819 s & \(0.85(4)\) & 2143 s & 2513 \\

      FG4       & 4 & 1 & 1 & 1 & --- & --- & --- & --- &
      \(0.95(1)\) & 2196 s & \(0.84(4)\) & 2462 s & 2931 \\
      \hline

      PQ3 / 4th & 5 & 3 & 1 & $-$ & \(0.1780(18)\) & \(0.1995(49)\) & \(0.1794(55)\) & --- 
      & \(0.83(2)(1)\) & 1513 s & \(0.82(4)\) & 1934 s & 2355 \\

      FG3       & 4 & 1 & 1 & $-$ & ---  & ---  & ---  & --- 
      & \(0.92(1)\) & 1816 s & \(0.82(5)\) & 2158 s & 2641 \\

      \hline
    \end{tabular}
  \end{center}
  \caption[tun40]{Tuning of a $40^4$ simulation. CPU times refer to runs utilizing 256 cores of the Iridis cluster.}
  \label{tun40}
\end{table*}

Despite the comparison done in the last section does not favour the
use of a force-gradient integrator, it is expected this integrator
will be of use for larger lattice volumes.  Thus we now consider a
thermalized $40^4$ simulation (using the same action parameters as
used in the previous sections) and proceeded with a similar tuning
analysis.  We show the results in Table \ref{tun40}.  Although we were
able to use a higher stepsize for the schemes which use a force-gradient 
integrator, the trajectory CPU times are still higher than
PQPQP runs.

Using the data available, one can compute ratios between the cost of
the FG schemes over PQPQP ones; we see they decrease for larger
volumes --- see Table \ref{volume}.  From these results, we can see
that at some increased volume (assuming constant physics) the 
force-gradient integrator will become more efficient than the PQPQP
integrator.

\begin{table}[htb]
 \begin{center}
   \begin{tabular}{|l|c|c|}
     \hline Nested scheme & FG4/PQ4 & FG3/PQ3 \\ $24^3\times32$ & 1.30 & 1.18 \\ $40^4$ &
     1.17 & 1.12 \\ \hline
   \end{tabular}
 \end{center}
 \caption{Cost ratios (FG over PQPQP).}
 \label{volume}
\end{table}

This cross-over point can be estimated from appealing to the
requirement that the equilibrium distribution must satisfy \(\langle
e^{\delta H}\rangle = 1\).  Expanding to second order, we have
\(\langle \delta H \rangle \sim \frac{1}{2} \langle \delta H^2
\rangle\), thus for second- and fourth-order integrators \(\langle
\delta H \rangle \sim \dt^4\) and \(\langle \delta H \rangle \sim
\dt^8\), respectively.  The cost has a trivial linear volume factor
and scales linearly with \(1/\dt\).  Since \(\delta H\) is extensive
in the volume we can equate it with the volume stepsize scaling, thus
we have \(\mbox{cost}^{\mbox{\tiny{PQPQP}}} \sim V^{5/4}\) and
\(\mbox{cost}^{\mbox{\tiny{FG}}}\sim V^{9/8}\), so the cost ratio will behave 
as $V^{-1/8}$.  Using the data in table \ref{volume} to estimate the 
multiplicative coefficient, we estimate
that the force-gradient integrator becomes more efficient at
\(V=56^4\) and \(V=50^4\) for the four-level and three-level
integrators, respectively.  Given current leading-edge lattice
computations are presently at this volume, or slightly larger, we
therefore predict that the force-gradient integrator will reduce
significantly the computational cost for future large-scale gauge
field ensemble generation.

\section{Conclusions}

We have presented a novel way of tuning a HMC integrator, together
with practical examples. This tuning procedure can be used for all
lattice gauge and fermionic actions, and allows a systematic study of
the MD integrators currently used in large scale dynamical lattice
simulations. We have also presented a new fourth-order integrator
which is expected to reduce significantly the computational cost of
HMC simulations.

We are currently working towards a general implementation of
the calculation of Poisson brackets and force-gradient terms 
in Chroma \cite{chroma}. Details about computing Poisson Brackets 
will be presented elsewhere \cite{prd}. In the near future we will 
also consider tuning simulations using other lattice actions.

\vspace*{0.15cm}

\begin{acknowledgments}
  PJS acknowledges support from FCT via grant SFRH/BPD/40998/2007, and
  project PTDC/FIS/100968/2008.  BJ acknowledges funding through US
  D.O.E Grants DE-FC02-06ER41440, DE-FC02-06ER41449 (SciDAC) and
  DE-AC05-060R23177 under which Jefferson Science Associates LLC
  manages and operates the Jefferson Lab.  MAC acknowledges support
  from the NSF via PHY-0835713 and OCI-1060067.  The U.S. Government
  retains a non-exclusive, paid-up, irrevocable, world-wide license to
  publish or reproduce this manuscript for U.S. Government purposes.
  The numerical results have been obtained using Chroma library
  ~\cite{chroma}.  Simulations have been carried out in Iridis
  (University of Southampton) and Centaurus (University of Coimbra)
  High Performance Computing facilities.
\end{acknowledgments}


\end{document}